\documentclass[conference]{IEEEtran}
\usepackage{algorithm,amsbsy,amsmath,amssymb,epsfig,bbm,mathrsfs,multirow,amsthm}
\usepackage{color}
\usepackage{algpseudocode,tabularx}
\usepackage{graphicx,caption,subcaption,array,multirow,bm}
\usepackage{makecell}
\usepackage{cite}
\usepackage{threeparttable}

\DeclareMathAlphabet{\mathpzc}{OT1}{pzc}{m}{it}

\makeatletter
\newcommand{\multiline}[1]{%
	\begin{tabularx}{\dimexpr\linewidth-\ALG@thistlm}[t]{@{}X@{}}
		#1
	\end{tabularx}
}
\makeatother
\begin{document}
\title{Deep Deterministic Policy Gradient for End-to-End Communication Systems without Prior Channel Knowledge}
\author{ Bolun Zhang and Nguyen Van Huynh\\
	School of Computing, Engineering and the Built Environment, Edinburgh Napier University, Edinburgh, UK
}
\maketitle
\begin{abstract}
End-to-End (E2E) learning-based concept has been recently introduced to jointly optimize both the transmitter and the receiver in wireless communication systems. Unfortunately, this E2E learning architecture requires a prior differentiable channel model to jointly train the deep neural networks (DNNs) at the transceivers, which is hardly obtained in practice. This paper aims to solve this issue by developing a deep deterministic policy gradient (DDPG)-based framework. In particular, the proposed solution uses the loss value of the receiver DNN as the reward to train the transmitter DNN. The simulation results then show that our proposed solution can jointly train the transmitter and the receiver without requiring the prior channel model. In addition, we demonstrate that the proposed DDPG-based solution can achieve better detection performance compared to the state-of-the-art solutions.
\end{abstract}


\section{Introduction}
\label{Sec:intro}

The existing communication system is designed based on multiple signal processing blocks where each is separately implemented and optimized for a particular task. This multi-block architecture has achieved decent performance, but at the cost of increasing design complexity \cite{8psk}. Besides, the transmitter and the receiver cannot be jointly optimized under this conventional architecture \cite{overAir}. Recently, deep learning (DL) technique has been applied to the communication system design, which interprets the E2E communication system as a DNN-based auto-encoder over an interference channel \cite{intro_phy}. This E2E-based learning method theoretically assumes the availability of the explicit channel model, which is perceived as an intermediate layer connecting the transmitter and the receiver. The auto-encoder architecture can therefore be trained in a supervised manner to jointly optimize for both the transmitter and the receiver. The biggest shortcoming of this E2E-based learning approach is that the training process is implemented based on the assumption of the prior differentiable channel model \cite{GradientFree}. However, the channel model in actual scenarios is usually considered as a black box, where the channel transfer function is typically non-differentiable and the channel gradients are difficult to estimate, especially in wireless communications. 

To address this problem, an alternating training approach is proposed in \cite{without}, where the authors perform supervised learning to train the receiver and reinforcement learning (RL) to train the transmitter. This approach demonstrated that the E2E communication systems can be trained without any prior assumption of channel model. In \cite{noisy_feedback}, an improved version of alternating training scheme with noisy feedback link is proposed. Although these methods can perform training without the knowledge of channel model, they can only work well with simple channel models, e.g., Additive White Gaussian Noise (AWGN) channel. Meanwhile, these solutions require long training time as it uses a ``Transformer'' network to estimate the channel responses, and a ``Discriminator'' network to recover the distorted signal.

In this paper, we propose a novel deep reinforcement learning (DRL) based E2E communication system using DDPG algorithm to address these challenges, where both the transmitter and the receiver can be trained over an unknown channel. Our major contributions are summarized as follows.
\begin{itemize}
    \item We develop a DDPG-based E2E communication system, which can jointly optimize the transmitter and the receiver without the prior knowledge of channel model.
    \item By utilizing convolutional neural network (CNN), our proposed scheme can achieve notable performance enhancement compared to the alternating training scheme for both the Rayleigh and Rician fading channels.
    \item Our solution can achieve a lower steady state of block error rate (BLER) compared to the alternating training scheme within the same training time.
\end{itemize}
\section{System Model}
\label{Sec.System}
\subsection{Auto-Encoder based End-to-End Communication Systems}
\begin{figure}[!]
  \centering
  \includegraphics[scale=0.3]{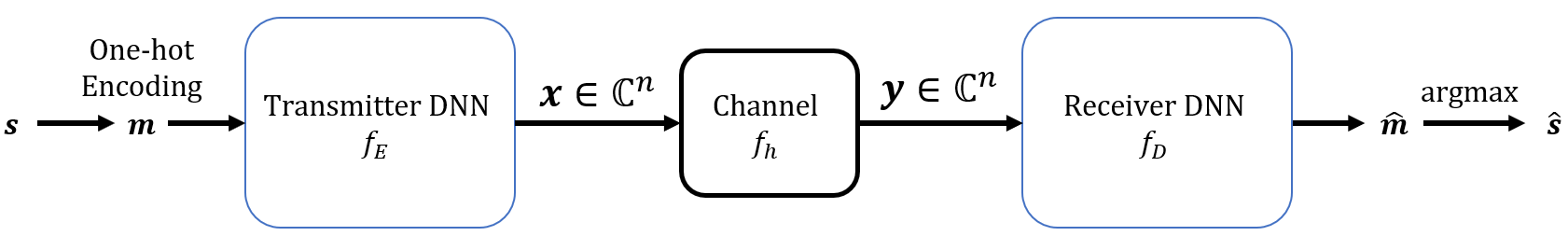}
  \caption{Overview of the typical E2E communication system.}
  \label{fig:E2E}
\end{figure}

We consider an auto-encoder based E2E communication system, as illustrated in Fig. \ref{fig:E2E}. The goal of E2E communication system is to jointly optimize the transceivers for better communication performance. In particular, it expresses the transmitter and the receiver as two independent DNNs such that the traditional signal processing blocks at the transmitter are represented as an encoder and the blocks at the receiver are represented as a decoder. The transmitter aims to reliably deliver symbol $\bm{s}$ to the receiver over the noisy channel, and the receiver aims to recover the distorted signal $\bm{y} \in {\mathbb{C}}^{n}$ to the original symbol $\bm{s}$. The symbol $\bm{s}$ is firstly converted to one-hot encoding vector $\bm{m} \in \mathbb{M}$, then the transmitter encodes the message $\bm{m}$ into the encoded signal $\bm{x} \in {\mathbb{C}}^{n}$, where ${\mathbb{C}}^{n}$ denotes that the encoded signal $\bm{x}$ is presented as complex numbers, making $n$ discrete uses of channel. The process of E2E communication system can be demonstrated as the cascade of three individual functions, which can be expressed as:
\begin{equation}
\hat{\bm{s}}=f_{D}(f_{h}(f_{E}(\bm{s}; \bm{\theta}_{E})); \bm{\theta}_{D}),
\label{eq:1}
\end{equation}
where $f_{E}$ represents the encoder function which maps the original symbol to the encoded signal, i.e., $\bm{x}=f_{E}(\bm{s}; \bm{\theta}_{E})$, with $\bm{\theta}_{E}$ denotes the trainable weights at the transmitter. $f_{h}$ represents the channel impairments with channel realization $\bm{h}$, i.e., $\bm{y}=f_{h}(\bm{x})$. $f_{D}$ represents the decoder function that aims to recover the received signal to the estimated symbol, i.e., $\hat{\bm{s}}=f_{D}(\bm{y}; \bm{\theta}_{D})$, and $\bm{\theta}_{D}$ denotes the trainable weights at the receiver. This E2E system is trained in an supervised manner which aims to minimize the loss function, i.e., $L=\mathcal{L}(\bm{m}, \hat{\bm{m}})$. The loss function $\mathcal{L}(\bm{s},\hat{\bm{s}})$ is regarded as the objective function which measures the distance between the original symbol $\bm{s}$ and the estimated message $\hat{\bm{s}}$ for such E2E framework.

Such an E2E learning approach regards the whole system as an auto-encoder and optimizes the DNNs for the transmitter and the receiver in a data-driven and supervised manner. The training process is implemented by computing the error gradients of the differentiable loss function with respect to the parameters at each layer of the DNNs. The calculated error gradient is then back-propagated through the network, allowing each layer to adjust its weights and biases to minimize the E2E loss. Therefore, the E2E learning approach requires to mathematically formulate a differentiable channel function in advance to allow the error gradients to be back-propagated from the receiver to the transmitter, hence obtaining the global optimization. 

\subsection{Limitations of End-to-End Communication Systems}
The primary shortcoming of the E2E paradigm is that it is built on the assumptions of the explicit and differentiable channel model, which may not be always practical in real-world scenarios. The channel effects in the actual communication systems are subject to various factors, such as signal attenuation, additive noise, multi-path fading, and time-varying channel realization. As a result, the actual channel models are often non-differentiable, making it challenging to optimize the system using conventional gradient-based method. The non-differentiable channel model potentially blocks the back-propagation process of the error gradients, which may result in local optimizations. This scenario can cause the system focus solely on optimizing the receiver network, neglecting the transmitter network and hindering the overall learning of the system. In addition, acquiring perfect CSI in practical scenarios can be challenging due to several factors, including hardware limitations and dynamic channel conditions \cite{raj2018backpropagating}. Consequently, it can be difficult to precisely model the channel behaviour, and the assumptions of the explicit and differentiable channel functions might limit or even impair the performance of E2E approach in practical communication scenarios. 

In this paper, we propose a DDPG-based E2E communication system to overcome the limitation of non-differentiable channel models and imperfect CSI. By employing experience replay technique \cite{ExperienceReplay}, the transmitter and the receiver can be jointly optimized without knowing any prior knowledge of the channel models. In the following sections, we will present our proposed DDPG solution in detail.

\section{Deep Deterministic Policy Gradient for E2E Communication Systems}
\label{sec:RL_system}
This section first provides the fundamentals of reinforcement learning and deep reinforcement learning. Then, the proposed DDPG-based framework for E2E communications is discussed.

\subsection{Deep Reinforcement Learning Basics}

A typical RL system consists of an agent and an environment, and the goal of RL is to train the agent to take actions in the unknown environment so as to collect the experience and maximize the cumulative reward by constant explorations. In particular, at time step $t$, the agent observes the environment, i.e., state $\bm{s}_{t}$, and makes action $\bm{a}_{t}$ based on its current policy $\pi$. After performing action $\bm{a}_{t}$, the agent observes the immediate reward $r_{t}$ and the next state $\bm{s}_{t+1}$. For a given policy $\pi$, the action-value function, i.e., Q-function, which measures the expected return for a given state-action pair, can be mathematically expressed as:
\begin{equation}
Q_{\pi}(\bm{s},\bm{a})=\mathbb{E}_{\pi}[G_{t}|\bm{s}_{t}=\bm{s},\bm{a}_{t}=\bm{a}],
\label{eq:Q}
\end{equation}
where $G_{t} = \sum^{\infty}_{t=1}\gamma^{t}r_{t}$ indicates the cumulative discounted reward for one trajectory, and $\gamma \in (0, 1]$ represents the discount factor which measures the importance of the long-term rewards to immediate rewards. In Q-learning, the update rule of the Q-value used for learning optimal policy can be expressed as:
\begin{equation}
Q(\bm{s},\bm{a}) {\leftarrow} Q(\bm{s},\bm{a})+{\alpha}[R(\bm{s},\bm{a})+{\gamma}\max Q(\bm{s}',\bm{a}')-Q(\bm{s},\bm{a})],
\end{equation}  
where $R(\bm{s},\bm{a})$ denotes the immediate reward given state $\bm{s}$ and action $\bm{a}$. $\bm{s}'$ denotes the next state that the agent transits to after performing action $\bm{a}$ in state $\bm{s}$, $\bm{a}'$ is the next action that the agent can take at state $\bm{s}'$, and $\alpha$ is the learning rate which determines how much weights is given to the new information. This algorithm iteratively improves the policy by updating the Q-values and selecting the action with the highest Q-value at each state, which aggressively select local optimum for every decision during exploration.

To further improve the convergence of Q-learning, deep Q-learning is proposed by replacing the Q-table by a DNN, namely deep Q-network (DQN) \cite{DQN}. It utilizes the power of DNN in solving regression problems, which takes the state-action pair as the input and approximates the corresponding Q-value by making predictions. The main advantage of deep Q-learning over Q-learning is the efficiency in handling complex tasks while Q-learning cannot obtain the optimal policy for complex problems in reasonable time. Another advantage of deep Q-learning is the experience replay method which stores the accumulated information in the buffer allowing the agent to sample from the past experience to optimize the policy network. The experience replay technique makes efficient use of the cumulative experience and solves the data correlation problem in the observation sequence by random sampling. However, the biggest shortcoming of deep Q-learning is that it cannot directly perform on continuous action space, while the action spaces in many real-world applications are continuous, e.g., the encoded signal in communication systems is time-varying and cannot be measured by discrete states. The second issue of deep Q-learning is that the high update frequency of the target network causes the behaviour network ``chasing'' a moving target. This paper introduces an advantaged deep Q-learning algorithm, namely Deep Deterministic Policy Gradient (DDPG), to not only handle the continuous action space but can also jointly train the transmitter and the receiver without prior channel information.

\subsection{End-to-End Communication Systems using DDPG}

DDPG is an off-policy actor-critic algorithm taking the advantages of deep Q-learning and policy gradient \cite{ddpg}. There are several significant features of the proposed DDPG-based system: (i) it uses deterministic policy for network optimizations, which allows the agent to directly operate on the continuous space action, (ii) it adopts actor-critic method, where the actor network is to generate the deterministic action for a given state and the critic network is to produce a score which measures the performance of the current action, and (iii) it trains two target networks respectively for the actor and the critic, and thus further increasing the training stability. The target networks are the time-delayed copies of their original networks that slowly track the behaviour networks, which significantly improves the stability in training process.

\begin{figure}[!]
  \centering
  \includegraphics[scale=0.29]{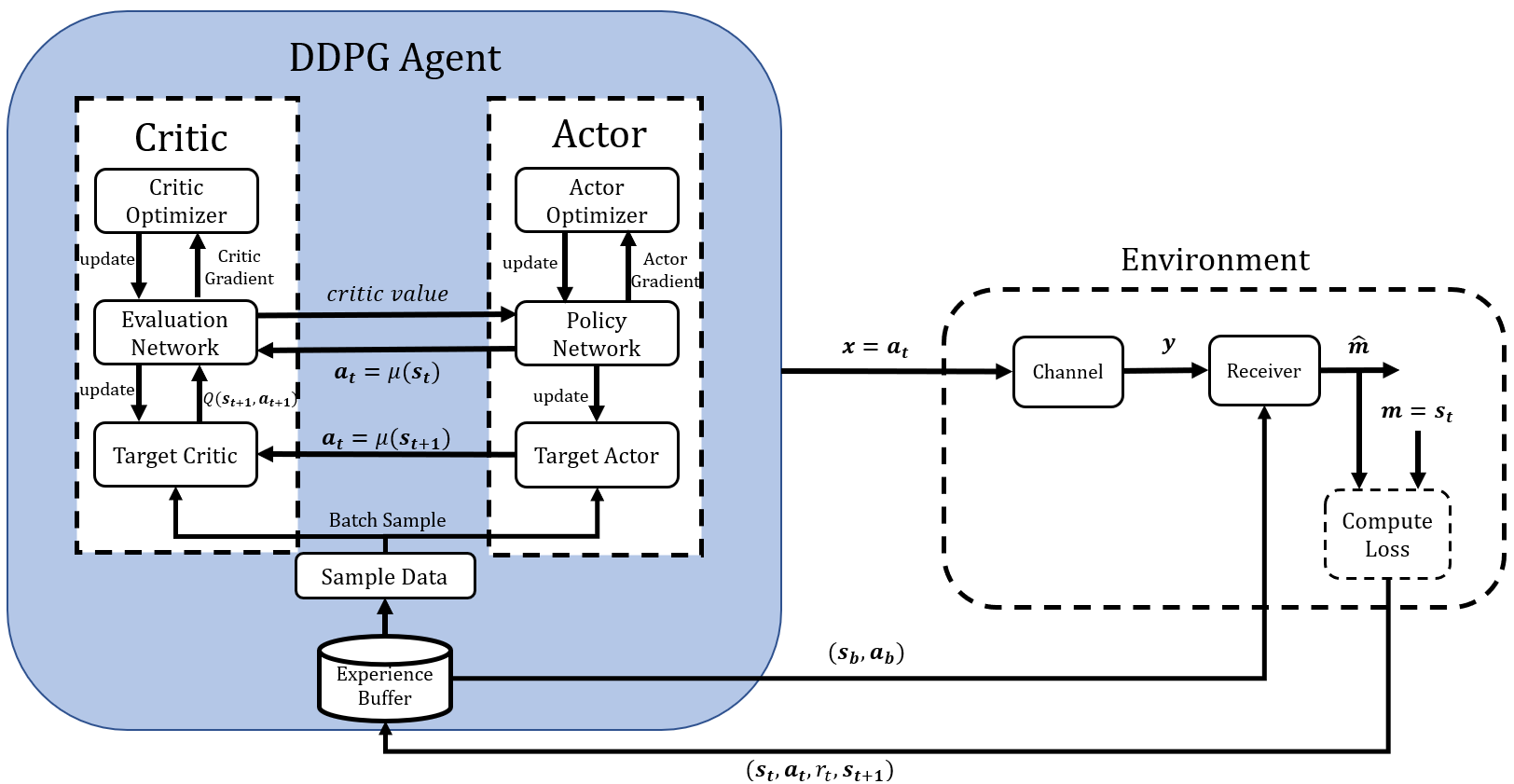}
  \caption{Overview of DDPG-based E2E communication system.}
  \label{fig:DDPG}
\end{figure}

An overview of the proposed DDPG-based E2E communication system is presented in Fig. \ref{fig:DDPG}, where the observation state is the input message, i.e., $\bm{s}_{t}=\bm{m}$. The action is the encoded signal, i.e., $\bm{a}_{t}=\bm{x}$. $\bm{y}$ is the received signal and $\hat{\bm{m}}$ is the estimated output message. $\bm{s_{b}}$ and $\bm{a_{b}}$ are the random state-action batches sampled from the experience buffer for training the receiver. The channel model and the receiver model are the major components of the environment. The DDPG agent consists of the actor and the critic networks, where the actor network, i.e., transmitter model, maps the input message $\bm{m}$ to the encoded signal $\bm{x}$ and sends it to the environment. The channel model in the environment distorts the encoded signal $\bm{x}$ and sends the damaged signal $\bm{y}$ to the receiver model which eventually decodes the damaged signal $\bm{y}$ to the estimated message $\hat{\bm{m}}$. The loss of the output is calculated using categorical cross-entropy loss. As mentioned, the loss of the receiver DNN will be fed back to the transmitter to update its DNNs. In particular, the reward at time $t$ after making an action at state $\bm{s}_{t}$ is the negative value of the computed loss, which can be expressed as follows:
\begin{equation}
    r_{t}=\frac{1}{n}\sum\limits_{i=1}^{n}{(\bm{s}_{t})}_{i} \cdot log[f_{\bm{\theta}_{R}}(\mu({(\bm{s}_{t})}_{i}\mid\bm{\theta}_{T}))],
\end{equation}
where $\bm{\theta}_{R}$ is the parameter of the receiver network, $\bm{\theta}_{T}$ is the parameter of the transmitter network, and $\mu(\bm{s}_{t}|\bm{\theta}_{T})$ is the action performed by the policy $\mu$ at the given state $s_{t}$ based on the transmitter parameters $\bm{\theta}_{T}$. $f_{\bm{\theta}_{R}}$ represents the process at the receiver model to map the damaged signal to the estimated output signal, and $n$ refers to the length of the input message.

The pseudocode of the whole training process is depicted in Algorithm 1. At the beginning stage, the actor and the critic networks are initialized with random parameters $\bm{\theta}^{\mu}$ and $\bm{\theta}^{Q}$, respectively. Two target networks ${\mu'}$ and ${Q'}$ are initialized with the same weights for initial balance. A replay buffer with capacity of $C$ is initialized to store the cumulative information. During the training stage, a random observation state, i.e., input message $\bm{s}_{t}=\bm{m}$ is generated from the environment at the beginning of each episode. For every time step in each episode, the actor network maps the current observation state $\bm{s}_{t}$ to the action, i.e., encoded signal $\bm{a}_{t}=\bm{x}$. The current observation state and action are sent to the environment, where the observation state is used as the label of the estimated output to calculate the loss value. The environment will then return the next observation state $\bm{s}_{t+1}$ and the calculated reward $r_{t}$ as feedback to the agent for updating the actor and critic networks. A collection of state, action, reward and next state will be stored in the experience buffer of the agent. After collecting the data, a mini-batch of experiences is randomly sampled from the replay buffer, and the target Q-value for each state-action pair in the mini-batch is computed using the following equation,
\begin{equation}
    \bm{y_{t}} = r_{t}+{\gamma}Q(\bm{s}_{t+1},\bm{a}_{t+1}).
\end{equation}

\begin{algorithm}[h]
	\caption{\strut DDPG for E2E Communication Systems}
	\label{alg:mainloop}
	\begin{algorithmic}[1]
		\State \multiline{Randomly initialize actor network ${\bf \mu}(\mathbf{s} \mid \bm{\theta}^\mu)$ and \\critic network $Q(\bm{s},\bm{a} \mid \bm{\theta}^Q)$}
		\State{Initialize target network ${Q}'$ and $\mathbf{\mu}'$ with the same weights}
		\State{Initialize replay buffer with capacity $C$ and batch size $B$}
		\For{\textit{episode = 1 to E}}
		\State{Initialize a random state for the input message  $\bm{s}_{t}=\bm{m}$}
		\State{Initialize the episodic reward as zero}
		\For{\textit{step = 1 to N}}
		\State{\multiline{Assign next state from last iteration to current state: $\bm{s}_{t}=\bm{s}_{t+1}$}}
		\State \multiline{Select the action (encoded signal) from actor network: $\bm{x} = \bm{a_t} = \mu(\bm{s} \mid \bm{\theta}^\mu)$}
		\State {Feed the state $\bm{{s}_t}$ and action $\bm{{a}_t}$ to the environment}
		\State \multiline{Observe the new state $\bm{s}_{t+1}$, reward $r_t$ and done information}
		\State \multiline{Store the transition $(\bm{s_t},\bm{a_t},r_t,\bm{s}_{t+1})$ to Buffer}
		\State \multiline{Sample a batch of $B$ transitions $(\bm{s_t},\bm{a_t},r_t,\bm{s}_{t+1})$}
		\State \multiline{Set $\bm{y_{t}}=r_{t}+\gamma{Q(\bm{s}_{t+1},\bm{a}_{t+1})}$}
		\State \multiline{Update critic network by minimizing the loss:}
		\begin{equation}
			L=\frac{1}{N}\sum\limits_{t}(\bm{y_{t}}-Q(\bm{s}_{t},\bm{a}_{t}|\bm{\theta}^{Q}))^{2} \nonumber
		\end{equation}
		\State \multiline{Update actor network using sampled policy gradient:}
		\begin{equation}
            {{\nabla }_{{{\bm{\theta} }^{\mu }}}}J\approx {\nabla}_{a}Q(\bm{s},\bm{a}){\nabla}_{{\bm{\theta}}^{\mu}}{\mu}(\bm{s}|{\bm{\theta}}^{\mu})	\nonumber		
		\end{equation}
		\State{Update target networks for both actor and critic:}
		\begin{equation}
			\begin{aligned}
				\bm{\theta}^{Q'}&=\tau\bm{\theta}^{Q}+(1-\tau)\bm{\theta}^{Q'} \\
				\bm{\theta}^{\mu'}&=\tau\bm{\theta}^{\mu}+(1-\tau)\bm{\theta}^{\mu'} \nonumber
			\end{aligned}
		\end{equation}
		\State \multiline{Sample a random batch of $D$ transitions $(\bm{a}_{t},\bm{s}_{t})$}
		\State \multiline{Train Receiver model with the sampled batch}
		\If{\textit{done==True}}
		\State{Break}
		\EndIf
		\EndFor
		\If{\textit{stop criterion met}}
		\State{Break Main Training Loop}
		\EndIf
		\EndFor
		\end{algorithmic}
	\end{algorithm}

During the optimization of the actor and the critic networks, we aim to maximize the expected return, i.e., the expected value of the Q-function $Q(\bm{s},\bm{a})$ by iteratively updating the actor and the critic networks. The expected value of the Q-function can be expressed as:
\begin{equation}
    J(\bm{\theta}) = \mathbb{E}[Q(\bm{s},\bm{a})|_{\bm{s}=\bm{s}_{t},\bm{a}_{t}=\mu(\bm{s}_{t})}],
\end{equation}
where $\bm{s}_{t}$ denotes the current state of the environment, and $\bm{a}_{t}$ denotes the current action selected by the policy function, i.e., $\bm{a}_{t}=\mu(\bm{s}_{t})$. The next essential step of the optimization of the actor network, i.e., the policy function $\bm{a}_{t}=\mu(\bm{s}_{t})$ is to take the gradient of the expected Q-value with respect to the policy parameters $\bm{\theta}_{\mu}$, which allows it to improve the expected cumulative reward. The gradient of the expected Q-value with respect to the policy parameter can be expressed as:
\begin{equation}
    {{\nabla }_{{{\bm{\theta} }^{\mu }}}}J\approx {\nabla}_{\bm{a}}Q(\bm{s},\bm{a}){\nabla}_{{\bm{\theta}}^{\mu}}{\mu}(\bm{s}|{\bm{\theta}}^{\mu}).
\end{equation}

After the optimization of the behaviour networks of the actor and the critic, we update their target networks based on soft update rule to slowly track the parameters of the learned networks. The optimization process of the target networks can be expressed as: 
\begin{equation}
    \bm{\theta}' = {\tau}{\bm{\theta}}+(1-\tau)\bm{\theta}',
\end{equation}
where $\bm{\theta}'$ denotes the parameters of target networks, $\bm{\theta}$ denotes the parameters of behaviour networks, and $\tau$ is a hyper-parameter called ``soft update rate'', which controls the blending between the behaviour network and target network weights. $\tau$ is defined as a positive value which is smaller than 1 for slow update, i.e., $\tau \ll 1$. The receiver model is trained in a supervised manner with the state-action pair batch randomly sampled from the buffer. In this way, the proposed solution can jointly optimize both the transmitter and the receiver without knowing the channel model in advance.

\subsection{Network Architecture of Transmitter and Receiver}
\begin{figure}[h!]
\begin{subfigure}{.26\textwidth}
  \centering
  \includegraphics[width=0.9\linewidth]{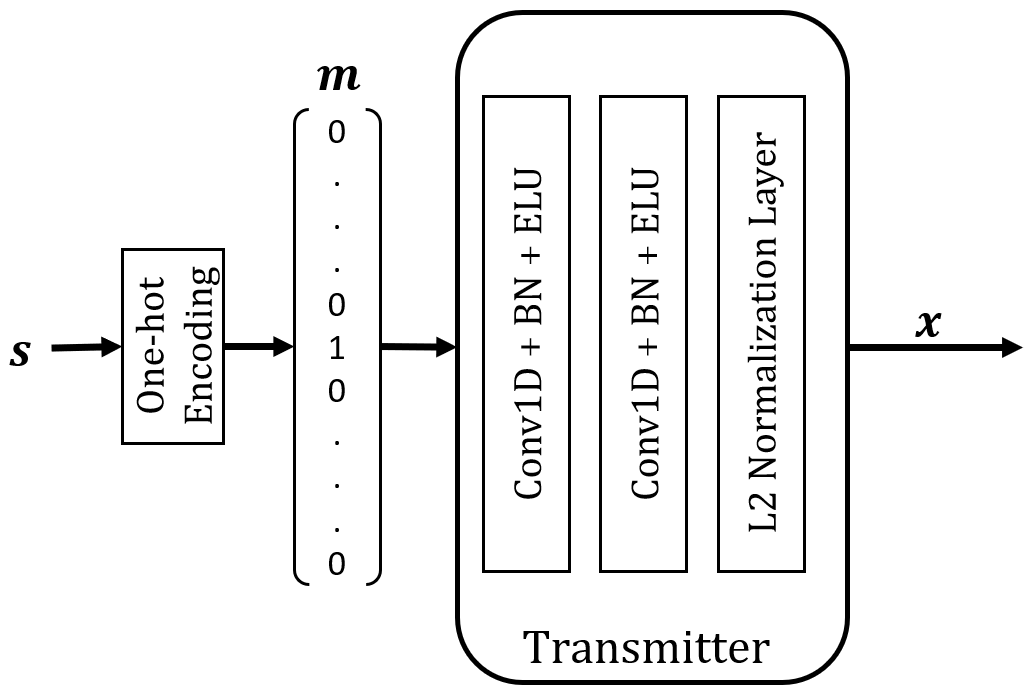}
  \caption{Transmitter}
  \label{fig:sfig3a}
\end{subfigure}%
\begin{subfigure}{.24\textwidth}
  \centering
  \includegraphics[width=0.9\linewidth]{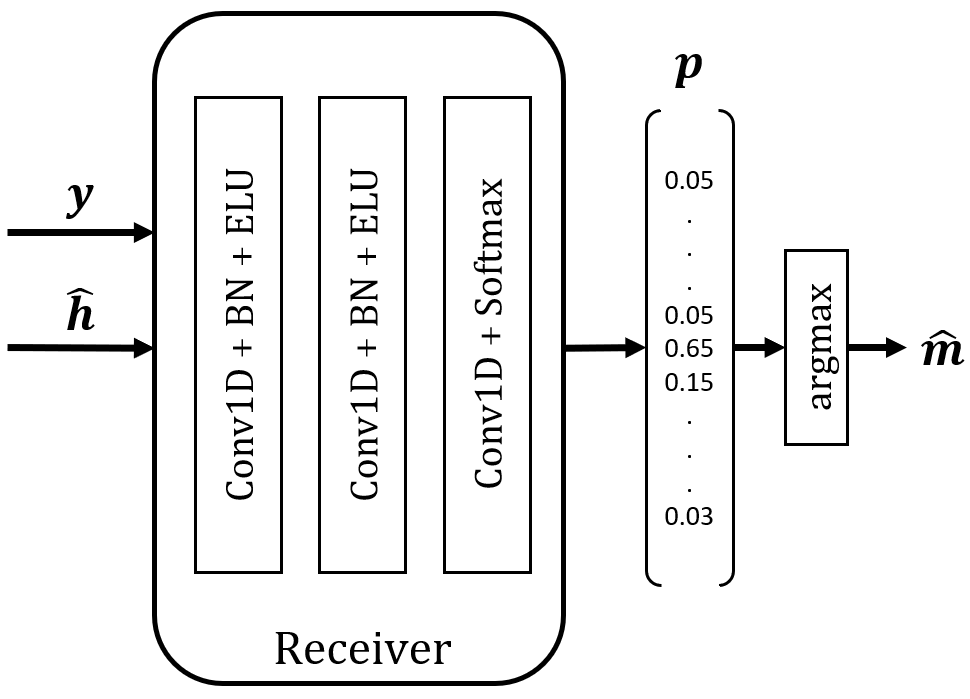}
  \caption{Receiver}
  \label{fig:sfig3b}
\end{subfigure}
\caption{Network architectures of (a) the transmitter and (b) the receiver.}
\label{fig:network_arch}
\end{figure}

The network architectures of the transmitter and the receiver are illustrated in Fig. \ref{fig:network_arch}. On the transmitter side, the symbol $\bm{s}$ is encoded to one-hot vector $\bm{m}$ of size $M$, i.e., $M=2^{k}$, where $k$ indicates the number of bits. The message $\bm{m}$ is fed to the transmitter. More specifically, the transmitter in Fig. \ref{fig:sfig3a} consists of two 1-Dimensional convolutional (Conv1D) layers, followed by an L2 normalization layer \cite{l2} which normalizes the total power of the encoded signal, i.e., $||\bm{x}||^{2} \leqslant n$. The first convolutional layer uses $M$ filters and is followed by a batch normalization layer and Exponential Linear Unit (ELU) activation function. The second convolutional layer uses $2n$ filters to reshape the length of the encoded signal. The encoded signal $\bm{x}$ is a one-dimensional vector with length of $2n$, representing the complex-valued symbol. For the receiver, the estimated channel response $\hat{\bm{h}}$ and the received signal $\bm{y}$ are together fed to the receiver network. The receiver consists of two Conv1D layers, followed by another Conv1D layer with Softmax activation function which produces a probability distribution vector of all possible output messages $\hat{\bm{m}}$. The algorithm then chooses the index with the highest probability to obtain the estimated symbol $\hat{\bm{s}}$.
\section{Simulation Results}
\label{sec:evaluation}
Several experiments have been implemented on Rayleigh and Rician fading channels to demonstrate the effectiveness of the proposed DDPG-based solution. The baseline for comparisons is the alternating training scheme proposed in \cite{without}. The following experiments are evaluated by BLER over different signal-to-noise ratio (SNR), ranging from 0 dB to 20 dB.   
\subsection{Parameter Setting}
For the hyper-parameters settings, the soft update rate $\tau$ is 0.005, which enables slow updates of the target networks. The learning rates of the actor and critic networks are set at 0.0002 and 0.0001, respectively. The numbers of training episodes and time step for every episode are set as 30,000 and 500, respectively. The adjacent states are uncorrelated due to the nature of the E2E communication system that every observation state, i.e., input message, is generated randomly. Therefore, the discount factor of Bellman function is 0.1. The training of the DDPG-based solution is implemented on Rayleigh and Rician fading channels, with SNRs of 20 dB and 10 dB. The size of the input message $M$ is set at 256, and the numbers of channel uses $n$ are set at 16 and 8 for different experiments. An additional pilot signal was leveraged for channel estimation. The Rician factor is set as 1 indicating the portions of the Line-of-sight components and None-line-of-sight components are equivalent.
\subsection{Performance Evaluation}
\begin{figure}[!]
  \centering
  \includegraphics[scale=0.2]{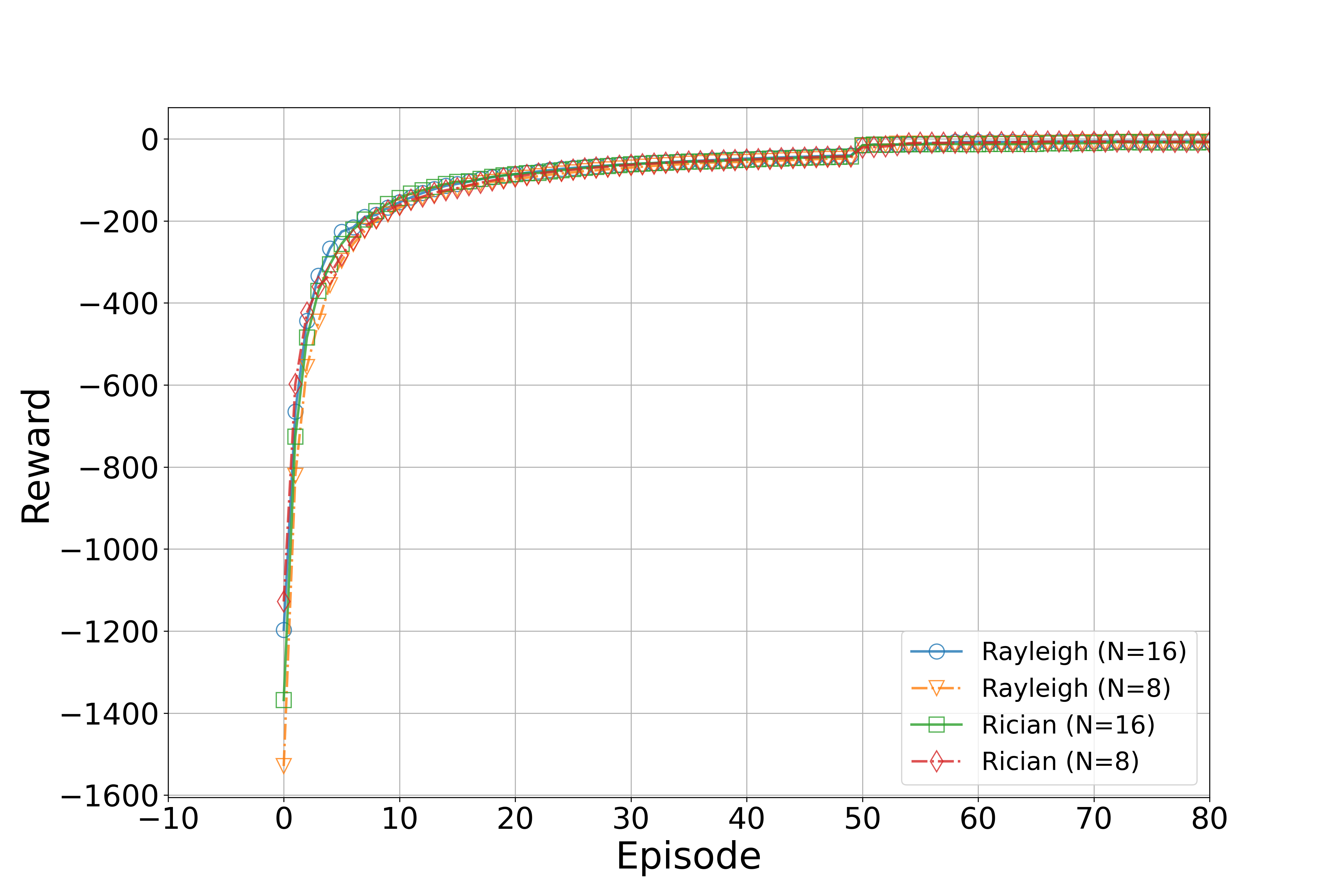}
  \caption{Episodic reward of Rayleigh and Rician fading channels.}
  \label{fig:reward}
\end{figure}

\begin{figure}[!]
  \centering
  \includegraphics[scale=0.2]{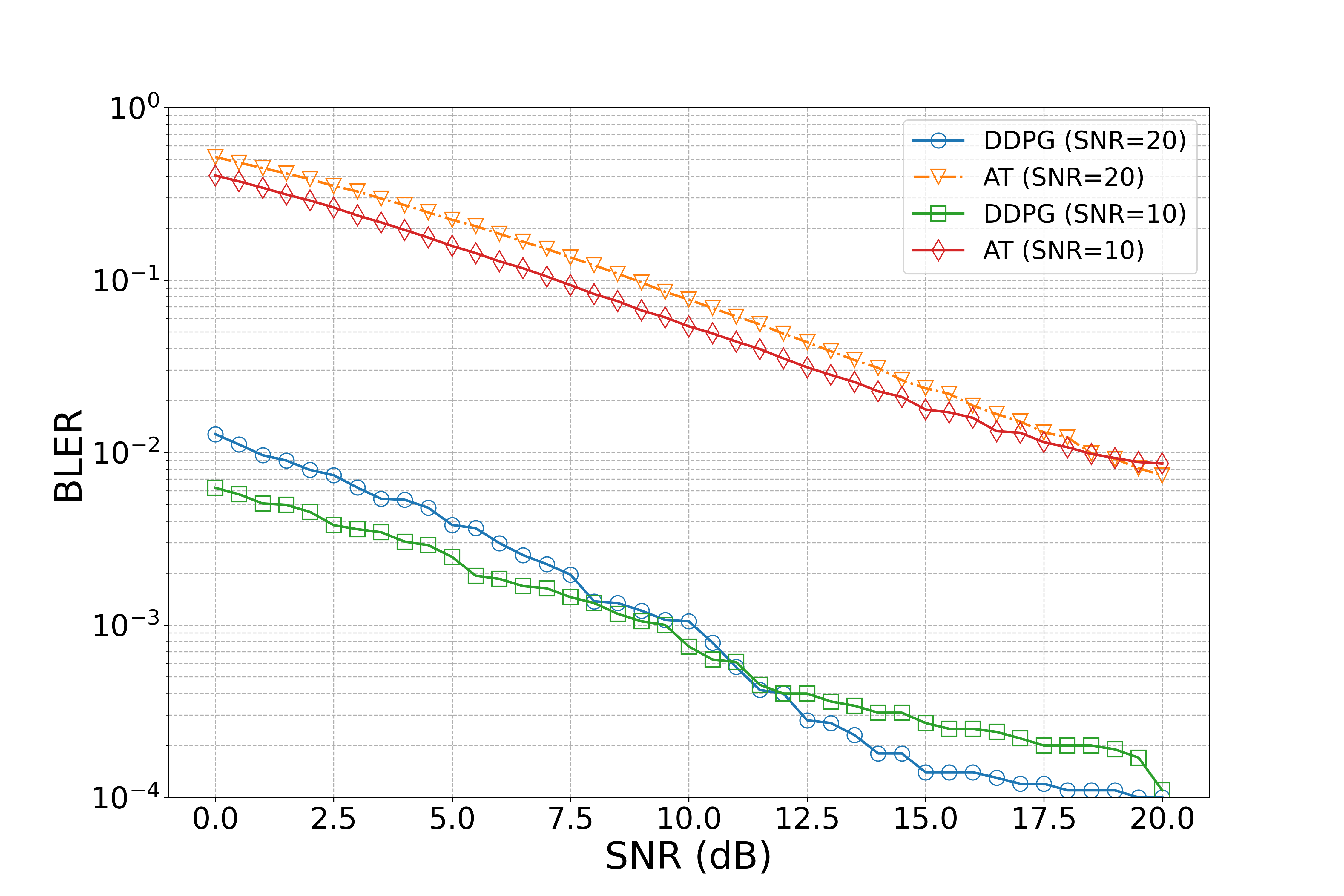}
  \caption{BLER of Rayleigh fading channel with channel size of 16.}
  \label{fig:Rayleigh16}
\end{figure}

\begin{figure}[!]
  \centering
  \includegraphics[scale=0.2]{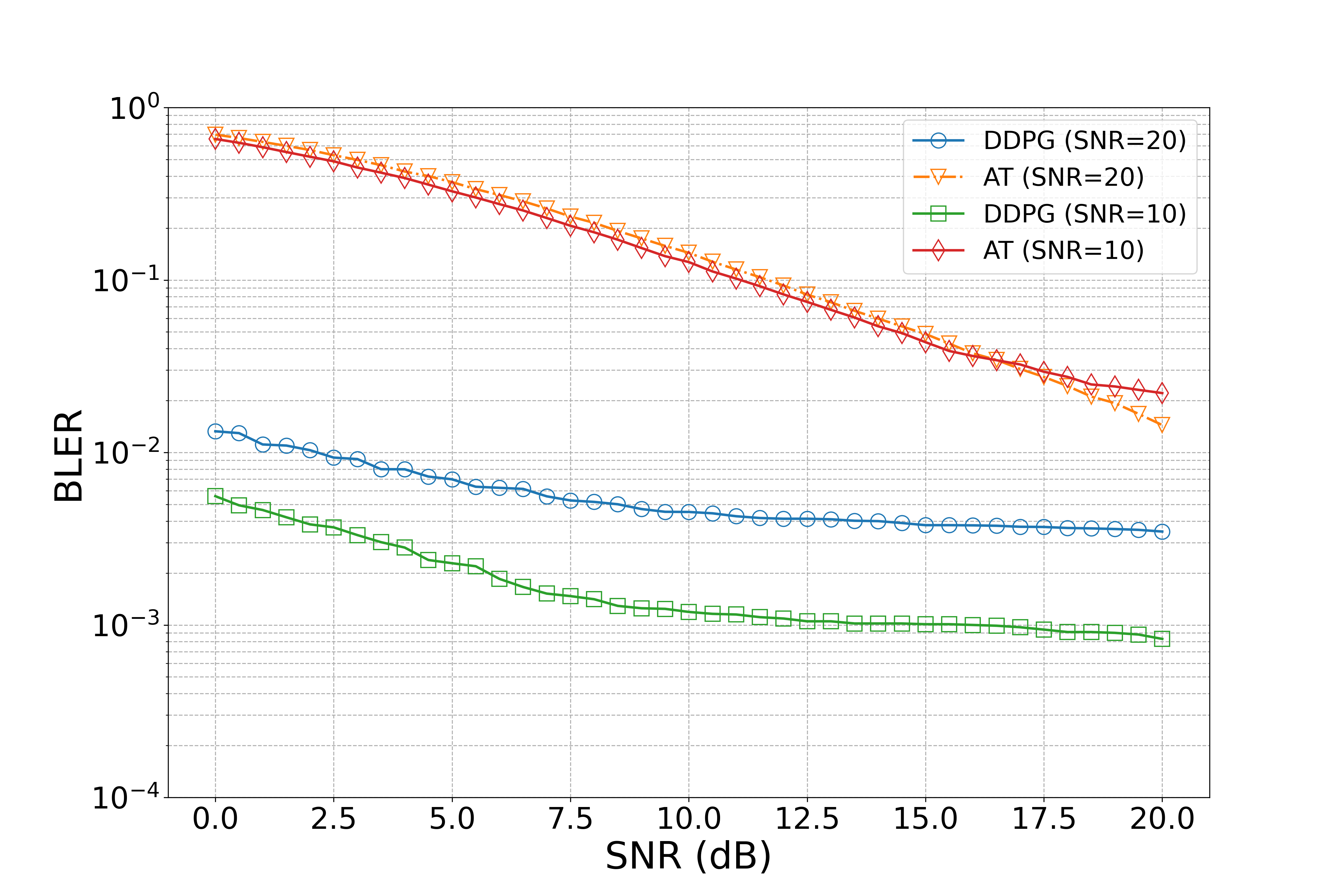}
  \caption{BLER of Rayleigh fading channel with channel size of 8.}
  \label{fig:Rayleigh8}
\end{figure}

\begin{figure}[!]
  \centering
  \includegraphics[scale=0.2]{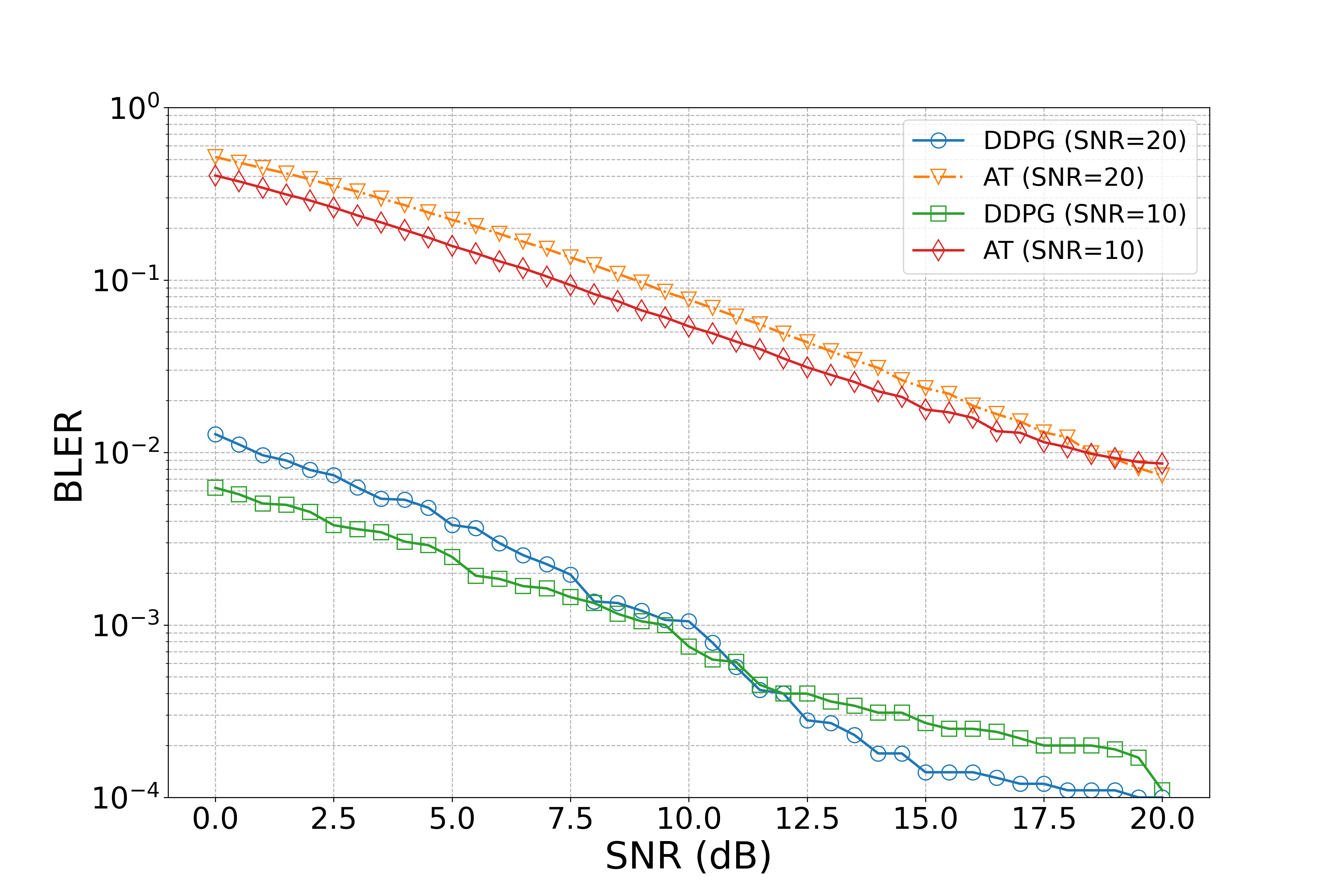}
  \caption{BLER of Rician fading channel with channel size of 16.}
  \label{fig:Rician16}
\end{figure}

\begin{figure}[!]
  \centering
  \includegraphics[scale=0.2]{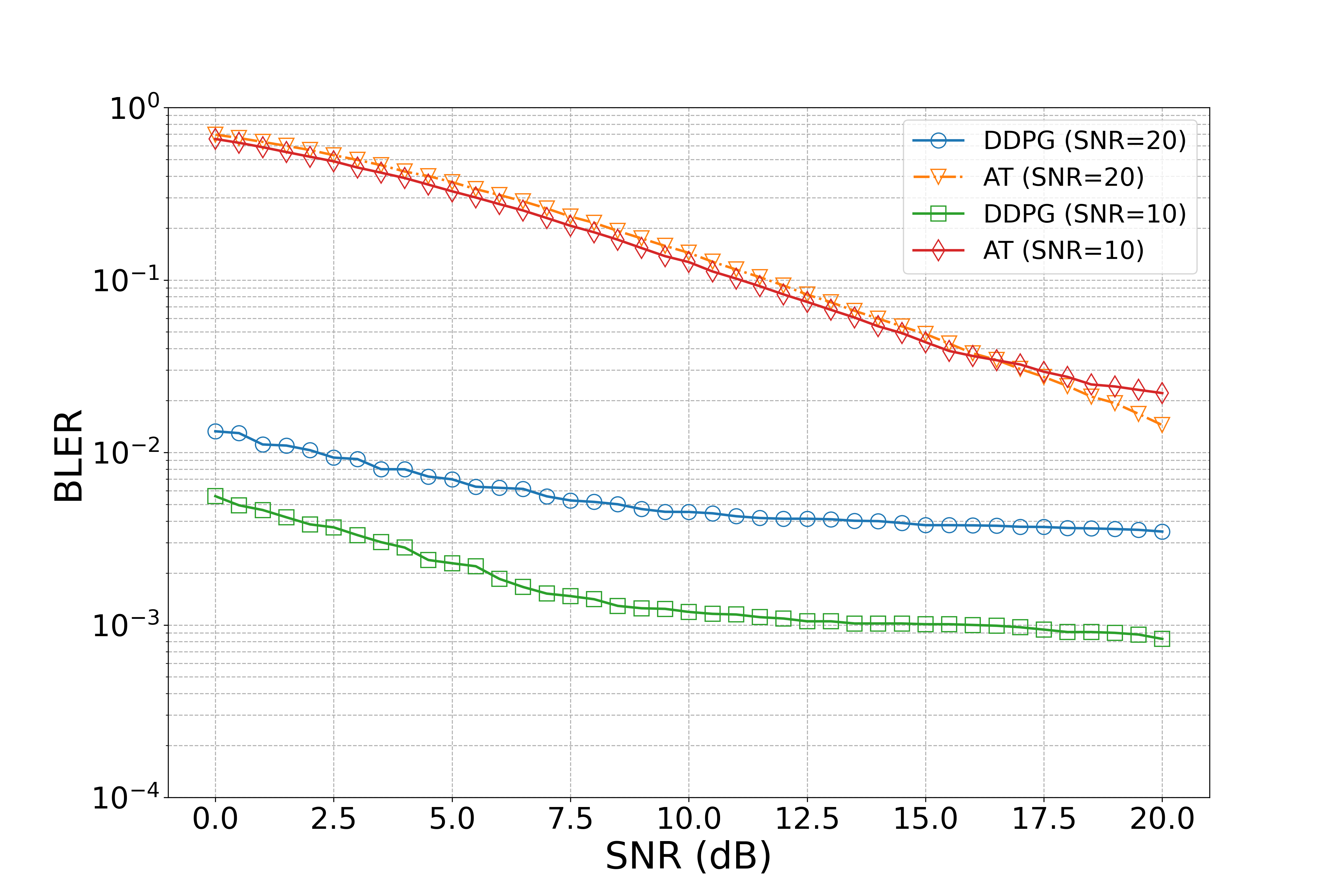}
  \caption{BLER of Rician fading channel with channel size of 8.}
  \label{fig:Rician8}
\end{figure}

The evolutions of the averaged reward over the first 80 episodes of both the Rayleigh and Rician fading channels with training SNR of 20 dB are presented in Fig. \ref{fig:reward}, averaged over the last 50 episodes. It suggests that the episodic reward of both fading channels can converge close to zeros within 80 episodes, indicating the feasibility of the proposed DDPG-based solution for training the E2E communication systems.
The following figures demonstrate the BLER performance of the proposed DDPG-based scheme against the alternating training scheme over Rayleigh and Rician fading channels. It can be observed that our proposed solution significantly outperforms the alternating training method. With regards to channel size $n$ of 16 under Rayleigh channel shown in Fig. \ref{fig:Rayleigh16}, the DDPG-based method with the training SNR of 10 dB shows better performance before 10 dB, but worse after 10 dB compared to that with the training SNR of 20 dB. This indicates that the model trained with SNR of 10 dB is able to detect the distorted signal with SNRs lower than 10 dB but difficult to detect the distorted signal with SNRs higher than 10 dB. The similar trend happens in the Rician fading channel with channel size of 16 as shown in Fig. \ref{fig:Rician16}. In particular, two BLER curves of the DDPG-based method, intersect at about 16 dB, indicating that the trained DDPG model with SNR of 10 dB outperforms that with SNR of 20 dB before the intersecting point. For channel size of 8, the proposed scheme with the training SNR of 10 dB shows better performance for the whole SNR range under both Rayleigh and Rician fading channels, as shown in Fig. \ref{fig:Rayleigh8} and Fig. \ref{fig:Rician8}. 

\begin{figure}[!]
\begin{center}
  \centering
  \includegraphics[scale=0.2]{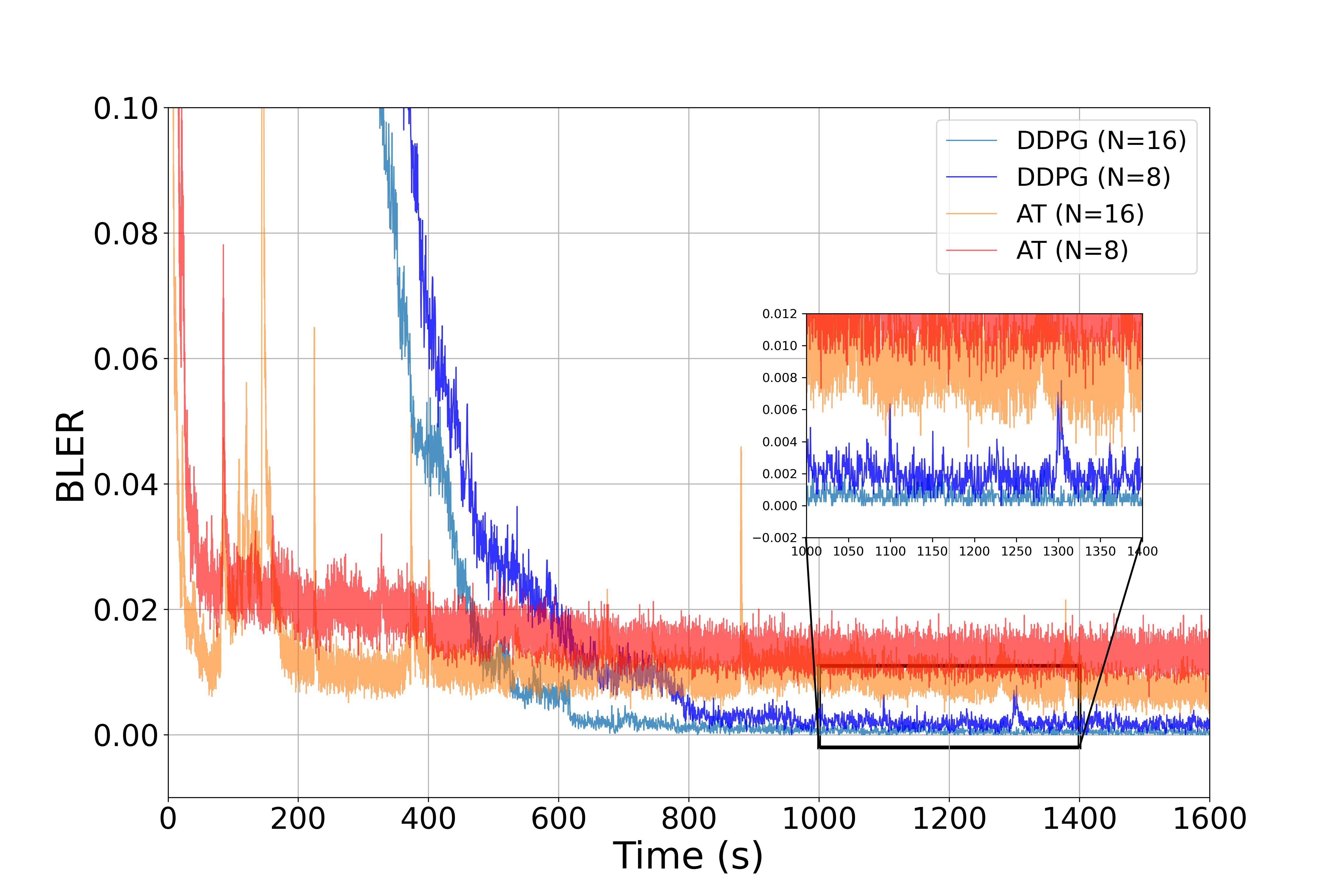}
  \caption{BLER vs. training time on Rayleigh fading channel.}
  \label{fig:BLERRayleigh}
\end{center}
\end{figure}

\begin{figure}[!]
\begin{center}
  \centering
  \includegraphics[scale=0.2]{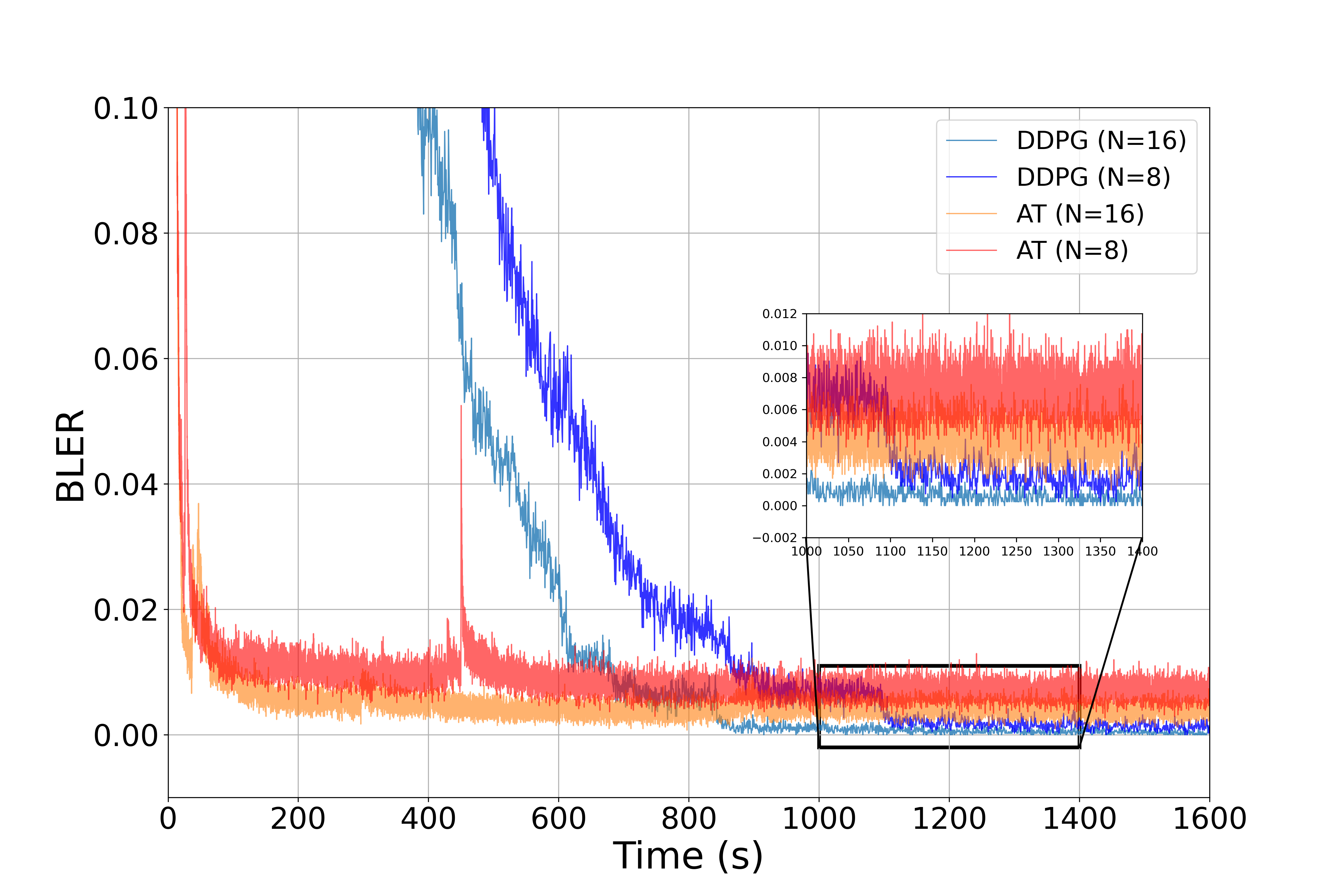}
  \caption{BLER vs. training time on Rician fading channel.}
  \label{fig:BLERRician}
\end{center}
\end{figure}

The BLER over training time of the two methods under different channel sizes are presented in Fig. \ref{fig:BLERRayleigh} and Fig. \ref{fig:BLERRician}, where AT refers to the alternating training scheme. The results show that the proposed DDPG-based solution can achieve lower steady state of BLER compared to the alternating training scheme. More specifically, as shown in Fig. \ref{fig:BLERRayleigh}, the proposed scheme can converge to the optimal policy within 800 seconds while the alternating training cannot converge after 1,600 seconds for Rayleigh fading channel. In Fig. \ref{fig:BLERRician}, the proposed scheme can converge within 1,000 seconds while the baseline scheme cannot converge after 1,600 seconds for Rician fading channel.

\section{Conclusion}
\label{sec:conclusion}
In this article, we have proposed a DDPG-based E2E learning solution to relax the requirement of the prior channel model. In particular, with the DDPG algorithm, the transmitter can update its DNN by learning from the reward, i.e., loss value, sent from the receiver, given the current state, i.e., bitstream, and the chosen action, i.e., encoded symbol. In this way, implicit information about the training process of the receiver DNN can be learned by the transmitter to adapt its DNN, and thus improving the whole system's performance. The simulation results have demonstrated the effectiveness of our proposed solution compared to existing solutions.

\end{document}